\documentclass[12pt]{article}
\usepackage{graphics}
\usepackage{amsmath}
\usepackage{amssymb}
\usepackage{hyperref}
\usepackage{graphicx}
\usepackage{makeidx}
\usepackage{mathrsfs}
\usepackage{amsfonts}
\usepackage[mathscr]{eucal}
\usepackage{amsmath}

\DeclareGraphicsExtensions{.eps,.pdf,.jpg,.png}

\def\RR{\vbox {\hbox to 8.9pt {I\hskip-2.1pt R\hfil}}}

\def\pni{\par\noindent}
\def\vsh{\smallskip}

\def\vsp{\vsh\pni} %% ie. \smallskip + \par

\numberwithin{equation}{section}

\begin{document}
%%%%%%%%%%%%%%%%%%%%%%%%%%%%%%%%%%%%%%%%%%%%%%%%%%%%%%%%%%%%%%%%%%%%%%%%
\font\title=cmbx12 scaled\magstep2
\font\bfs=cmbx12 scaled\magstep1
\font\little=cmr10
\begin{center}
{\title Energy dissipation in linear viscoelasticity}
\\ [0.50 truecm]
Juan Luis GONZ{\'A}LEZ  SANTANDER$^{(1)}$,
 %\\  [0.20 truecm]
 Francesco MAINARDI$^{(2)}$
 \\[0.50 truecm]
 $^{(1)}$  Department of Mathematics, University of  Oviedo.
 \\ C/ Leopoldo Calvo Sotelo 18, 33007 Oviedo, Spain;
 {gonzalezmarjuan@uniovi.es}
 \\
$^{(2)}$
 Department of Physics and Astronomy, University of Bologna and INFN.
\\ {Via Irnerio 46, I-40126 Bologna, Italy};
\\{ francesco.mainardi@unibo.it; mainardi@bo.infn.it; fracalmo@gmail.com}

%%
%\vskip 0.25truecm %% [0.25truecm]
%{\bf Version of \today}
 \vskip 0.20truecm
 % [0.25truecm]
 {\bf Published in Mathematics (MDPI)}
  \\
  {\bf Vol.14 No 16 (2026), 2936/1--18;  DOI: 10.3390/math14162936}
\\
{\bf SI: Special Functions: Representations and Applications}
\end{center}

%%%%%%%%%%%%
\begin{abstract}
We compare the specific dissipation $Q^{-1}\left( \omega \right) $ for the
Becker, Lomnitz, and Lambert models in linear viscoelasticity.
Graphically, the specific dissipation of these models behaves similarly as $\omega \rightarrow 0^{+}$ and $\omega \rightarrow +\infty$.
Furthermore, we obtain analytic formulas for the asymptotic
behavior of $Q^{-1}\left( \omega \right) $ in the Becker and Lomnitz models as $\omega \rightarrow 0^{+}$ and $\omega \rightarrow +\infty$,
where the latter limit coincides with the specific dissipation of the Maxwell model.
We also derive a novel closed-form expression for the specific dissipation in the
generalized Becker model, $Q_{\nu }^{-1}\left( \omega \right) $,  when the
parameter $\nu \in \left( 0,1\right] $ is rational. When $\nu \rightarrow 0$,
we recover the specific dissipation of the Maxwell model, whereas when $\nu =1$,
we recover that of the original Becker model. In addition, we
derive two new closed-form expressions for the specific dissipation
for the extended Jeffreys--Lomnitz model $Q_{\alpha }^{-1}\left( \omega
\right) $, the first being valid for $\alpha \in \left( 0,1\right] $,
and the second for $\alpha \in \left( -\infty ,1\right] $. When $\alpha
\rightarrow 0$, we recover the specific dissipation of the Lomnitz model,
whereas when $\alpha =1$, we recover that of the Maxwell model. Finally,
we conclude that the asymptotic behavior of $Q_{\alpha }^{-1}\left( \omega
\right) $ as $\omega \rightarrow +\infty $ coincides with the specific dissipation of the Maxwell model.

 \vsp {\bf Keywords}:
{Linear viscoelasticity; specific dissipation; generalized Becker model;
extended Jeffreys-Lomnitz model}

\vsp {\bf MSC}: {74-10; 33C20; 33E12; 33B20; 33B99; 44A10; 42A38}

\end{abstract}
\newpage

%
% \maketitle

%%%%%%%%%%%%%%%%%%%%%%%%%%%%%
%\begin{document}

%%%%%%%%%%%%%%%%%%%%%%%%%%%%%%%%%%%%%%%%%%
%\setcounter{section}{-1} %% Remove this when starting to work on the template.
%\section{How to Use this Template}

\section{Introduction}

In this section, we introduce the fundamentals of linear viscoelasticity.
First, we introduce the material functions $J\left( t\right) $ and $G\left(
t\right) $, as well as the dynamic functions $J^{\ast }\left( \omega\right) $
and $G^{\ast }\left( \omega\right) $, to obtain expressions for the specific
dissipation function $Q^{-1}\left( \omega \right) $. These concepts are
discussed in most treatises on linear viscoelasticity (see, for instance,
\cite{GrossBOOK53,PipkinBOOK86,TschoegelBOOK89,MainardiBook}). %MDPI: We revised the reference citation format and reordered all references in the whole paper to keep them in numerical order. Please confirm them all AUTHOR -> “I confirm”

Second, we introduce the classical models in Earth rheology, i.e., the Becker
and Lomnitz models, together with a recent model based on the Lambert $W$
function \cite{LambertJL}. For the history and further details of the Becker
and Lomnitz models, the reader is referred to the book by
 Mainardi \cite{MainardiBook}.
%and to several papers co-authored by him, (see, for instance %\cite{MainardiLomnitz} and \cite{BeckerLomnitzLambert}).
We also introduce generalized versions of the classical Becker and Lomnitz
models (see \cite{MainardiBecker,JeffreysLomnitz,MainardiLomnitz}).

\subsection{The Material Functions}

To characterize the response of a viscoelastic material subjected to an
applied stress $\sigma \left( t\right) $ or strain $\epsilon \left( t\right)
$, a unit-step excitation $\theta \left( t\right) $ is usually used, where $%
\theta \left( t\right) $ denotes the Heaviside function and $t\geq 0$ is
time. In this context, the \textit{creep compliance} %MDPI: Please confirm if the italics is unnecessary and can be removed. The following highlights are the same. 
% AUTHOR -> “It should be italic”
 $J\left( t\right) $ is
defined as the strain response to a unit-step of stress, and the \textit{%
relaxation modulus} $G\left( t\right) $ is defined as the stress response to
a unit-step of strain  (\cite{MainardiBook}, Equation (2.1)): %MDPI: Please check whether this refers to the Equation of this paper, if so, please revise the Equation number; same as below. 1. Please carefully check to make sure there are no Duplicate Equations throughout the text, thank you.
%2. The italics of variables with the same meaning should be consistent in the full text. Please check the full text and modify.

\vspace{-4pt}
\begin{eqnarray}
&&\sigma \left( t\right) =\theta \left( t\right) \quad \Rightarrow \quad
\epsilon \left( t\right) =J\left( t\right) , \\
&&\epsilon \left( t\right) =\theta \left( t\right) \quad \Rightarrow \quad
\sigma \left( t\right) =G\left( t\right) .
\end{eqnarray}%
The functions $J\left( t\right) $ and $G\left( t\right) $ are usually
referred to as the \textit{material functions}. In Earth rheology, the creep
compliance is usually written (\cite{MainardiBook}, Equation (2.80)):%
\begin{equation}
J\left( t\right) =J_{U}\,\left[ 1+q\,\psi \left( t\right) \right] ,\quad
t\geq 0,
\end{equation}%
where $J_{U}$ is the unrelaxed compliance, $q$ is a positive dimensionless
material constant, and $\psi \left( t\right) $ is the dimensionless creep
function. Note that the scaling factor $q$ accounts for differences among
materials following the same creep model, i.e., the same dimensionless creep
function $\psi \left( t\right) $. From now on, we set $q=1$. Moreover, since $J_{U}=J\left( 0\right) $,
we have%
\begin{equation}
\psi \left( 0\right) =0.  \label{psi(0)=0}
\end{equation}%
The dimensionless relaxation modulus is defined
 as in \cite{MainardiBook}, Equation (2.88),
 % {by Mainardi}
\begin{equation}
\phi \left( t\right) =J_{U}\,G\left( t\right) .  \label{phi(t)_def}
\end{equation}%
Note that $J_{U}$ acts as a scaling factor for the dimensionless relaxation
modulus $\phi \left( t\right) $. This dimensionless relaxation modulus obeys
the Volterra integral equation, see (\cite{MainardiBook}, Equation (2.89)) taking $q=1$:%
\begin{equation}
\phi \left( t\right) =1-\int_{0}^{t}\psi ^{\prime }\left( \tau \right)
\,\phi \left( t-\tau \right) \,d\tau .
\end{equation}%
Now, let the Laplace transform of a given function $f\left( t\right) $
(\cite{Gradshteyn}, Section 17.11): %MDPI:  Please check whether this refers to the Section of this paper, if so, please revise the Section number; same as below

\begin{equation}
\tilde{f}\left( s\right) :=\mathcal{L}\left[ f\left( t\right) ;s\right]
:=\int_{0}^{\infty }f\left( t\right) e^{-st}dt.  \label{Laplace_def}
\end{equation}%
The Laplace transform of the dimensionless relaxation
modulus is \cite{BeckerLomnitzLambert}:%
\begin{equation}
\mathcal{L}\left[ \phi \left( t\right) ;s\right] =\frac{1}{s\left[ 1+%
\mathcal{L}\left[ \psi ^{\prime }\left( t\right) ;s\right] \right] }.
\label{L[phi(t)]}
\end{equation}
\subsection{The Dynamic Functions and the Specific \\ Dissipation Function}

Another widely used form of excitation in viscoelasticity is the sinusoidal
excitation, which is (ideally) assumed to have been applied since $%
t\rightarrow -\infty $. In this context, we define the complex compliance $%
J^{\ast }\left( \omega\right) $ and the complex modulus $G^{\ast }\left(
\omega\right) $:
\begin{eqnarray}
&&\sigma \left( t\right) =e^{i\omega t} \quad \Rightarrow \quad \epsilon
\left( t\right) =J^{\ast }\left( \omega\right) e^{i\omega t}, \\
&&\epsilon \left( t\right) = e^{i\omega t} \quad \Rightarrow \quad \sigma
\left( t\right) =G^{\ast }\left( \omega\right) e^{i\omega t},
\end{eqnarray}%
where $\omega >0$ denotes the angular frequency of the excitation. The
material responses, $J^{\ast }\left( \omega\right) $ and $G^{\ast }\left(
\omega\right) $, are usually referred to as the \textit{dynamic functions}.
Together with the material functions $J\left( t\right) $ and $G\left(
t\right) $, they provide a complete description of viscoelastic
behavior. The relation between the relaxation modulus $G\left( t\right) $\
and the complex modulus $G^{\ast }\left( \omega\right) $ is given by (\cite{MainardiBook}, Equation (2.46)):%
\begin{equation}
G^{\ast }\left( \omega \right) =\left. s\,\tilde{G}\left( s\right)
\right\vert _{s=i\,\omega }.  \label{complex_modulus_def}
\end{equation}

The dynamic experiments provide information about storage and dissipation of
mechanical energy in the viscoelastic material. Usually, the energy
dissipation in a viscoelastic medium is measured by introducing the
so-called \textit{specific dissipation}
function (or \textit{internal
friction}), denoted as $Q^{-1}\left( \omega \right) $. According to (\cite{MainardiBook}, Equation (2.63))%
\begin{equation}
Q^{-1}\left( \omega \right) =\frac{\Im \left[ G^{\ast }\left( \omega \right) %
\right] }{\Re \left[ G^{\ast }\left( \omega \right) \right] }.
\label{Loss_tangent_def}
\end{equation}%
Note that, according to (\ref{phi(t)_def}), (\ref{Laplace_def}), and (\ref%
{L[phi(t)]}), we have%
\begin{equation}
s\,\tilde{G}\left( s\right) =s\,\mathcal{L}\left[ G\left( t\right) ;s\right]
=\frac{s}{J_{U}}\mathcal{L}\left[ \phi \left( t\right) ;s\right] =\frac{1}{%
J_{U}\left[ 1+\mathcal{L}\left[ \psi ^{\prime }\left( t\right) ;s\right] %
\right] },  \label{s*L[G]}
\end{equation}%
thus, according to (\ref{complex_modulus_def}) and (\ref{s*L[G]}),
\begin{equation}
G^{\ast }\left( \omega \right) =\frac{1}{J_{U}\left[ 1+\mathcal{L}\left[
\psi ^{\prime }\left( t\right) ;i\,\omega \right] \right] }.
\end{equation}%
However, for $z\in
%TCIMACRO{\U{2102} }%
%BeginExpansion
\mathbb{C}
%EndExpansion
$, we have%
\begin{equation}
\frac{\Im \left( \frac{1}{z}\right) }{\Re \left( \frac{1}{z}\right) }=-\frac{%
\Im \left( z\right) }{\Re \left( z\right) },
\end{equation}%
hence, we obtain the following expression for the specific dissipation
function:
\begin{equation}
Q^{-1}\left( \omega \right) =-\frac{\Im \left( \mathcal{L}\left[ \psi
^{\prime }\left( t\right) ;i\,\omega \right] \right) }{1+\Re \left( \mathcal{%
L}\left[ \psi ^{\prime }\left( t\right) ;i\,\omega \right] \right) }.
\label{Q-1_resultado_general}
\end{equation}
It is worth noting that throughout this article, we consider the principal branch of any multivalued functions on the real line or complex plane.

To obtain an alternative expression for the specific dissipation function,
we define the sine and cosine Fourier transforms  (\cite{Gradshteyn}, Section 17.31):%
\begin{eqnarray}
\mathcal{F}_{S}\left[ f\left( t\right) ;\omega \right] &:=&\sqrt{\frac{2}{%
\pi }}\int_{0}^{\infty }f\left( t\right) \sin \omega t\,\,dt,
\label{Fourier_sine_def} \\
\mathcal{F}_{C}\left[ f\left( t\right) ;\omega \right] &:=&\sqrt{\frac{2}{%
\pi }}\int_{0}^{\infty }f\left( t\right) \cos \omega t\,dt.
\label{Fourier_cosine_def}
\end{eqnarray}%
Applying the derivative theorem of the Laplace transform (\cite{Schiff}, Theorem 2.7) %MDPI:  Please check whether this refers to the Theorem of this paper, if so, please revise the Theorem number; same as below

\begin{equation}
\mathcal{L}\left[ f^{\prime }\left( t\right) ;s\right] =s\mathcal{L}\left[
f\left( t\right) ;s\right] -f\left( 0^{+}\right) ,
\end{equation}%
and taking into account (\ref{psi(0)=0}), we can rewrite (\ref%
{Q-1_resultado_general})\ as%
\begin{equation}
Q^{-1}\left( \omega \right) =-\frac{\Im \left( i\omega \,\mathcal{L}\left[
\psi \left( t\right) ;i\,\omega \right] \right) }{1+\Re \left( i\omega \,%
\mathcal{L}\left[ \psi \left( t\right) ;i\,\omega \right] \right) }.
\end{equation}%
Using the following identities for $z\in
%TCIMACRO{\U{2102} }%
%BeginExpansion
\mathbb{C}
%EndExpansion
$,%
\begin{eqnarray}
\Re \left( i\,z\right) &=&-\Im \left( z\right) , \\
\Im \left( i\,z\right) &=&\Re \left( z\right) ,
\end{eqnarray}%
we obtain
\begin{equation}
Q^{-1}\left( \omega \right) =\frac{\omega \,\Re \left( \mathcal{L}\left[
\psi \left( t\right) ;i\,\omega \right] \right) }{\omega \,\Im \left(
\mathcal{L}\left[ \psi \left( t\right) ;i\,\omega \right] \right) -1}.
\end{equation}%
Furthermore, according to the definitions of the Laplace transform (\ref%
{Laplace_def}), and the sine and cosine Fourier transform (\ref%
{Fourier_sine_def}) and (\ref{Fourier_cosine_def}), we have
\begin{eqnarray}
\mathcal{L}\left[ \psi \left( t\right) ;i\,\omega \right] &=&\int_{0}^{%
\infty }e^{-i\omega t}\,\psi \left( t\right) \,dt \\
&=&\int_{0}^{\infty }\cos \omega t\,\psi \left( t\right)
\,dt-i\int_{0}^{\infty }\sin \omega t\,\psi \left( t\right) \,dt \\
&=&\underset{\Re \left( \mathcal{L}\left[ \psi \left( t\right) ;i\,\omega %
\right] \right) }{\underbrace{\sqrt{\frac{\pi }{2}}\mathcal{F}_{C}\left[
\psi \left( t\right) ;\omega \right] }}-i\underset{-\Im \left( \mathcal{L}%
\left[ \psi \left( t\right) ;i\,\omega \right] \right) }{\,\underbrace{\sqrt{%
\frac{\pi }{2}}\mathcal{F}_{S}\left[ \psi \left( t\right) ;\omega \right] }}.
\end{eqnarray}%
Thus, we obtain the following expression for the specific dissipation
function:%
\begin{equation}
Q^{-1}\left( \omega \right) =-\frac{\omega \,\mathcal{F}_{C}\left[ \psi
\left( t\right) ;\omega \right] }{\omega \,\mathcal{F}_{S}\left[ \psi \left(
t\right) ;\omega \right] +\sqrt{\frac{2}{\pi }}}.
\label{Q-1_resultado_general_2}
\end{equation}

\subsection{Models in Linear Viscoelasticity}

Among the models in linear viscoelasticity compatible with the second law of
thermodynamics are those whose relaxation moduli $G\left( t\right) $ are
represented by completely monotone (CM) functions of time. We recall that a
CM function $f(t)$ is a non-negative, infinitely differentiable function
with derivatives alternating in sign for $t>0$, whereas a Bernstein function
is a non-negative function whose derivative is CM. A necessary and
sufficient condition for a function to be CM is provided by the Bernstein
theorem, which states that $f(t)$ is the Laplace transform of a non-negative
real function (see, for example, \cite{HanygaRHEOACTA05}, as reported in
Mainardi's book (\cite{MainardiBook}, Section 2.7)). The CM property was
formerly imposed indirectly in \cite{GrossBOOK53}, which is physically
equivalent to discrete or continuous, non-negative distributions of simple
exponential relaxations. In this respect, the determination of the specific
dissipation function $Q^{-1}\left( \omega \right) $ is relevant for all
models in linear viscoelasticity based on this CM property.

Since this determination is well known in the elementary mechanical models
(represented by springs and dashpots) including their generalized versions
obtained by adopting the operators of fractional calculus, we devote this
study to other models that find application mainly in Earth rheology. In
Section \ref{Section: classical models}, we start with the classical models
by Becker and Lomnitz, together with the more recent model based on the
Lambert $W$ function, in order to present a comparative view of the specific
dissipation in these models. In Section \ref{Section: Becker generalization}%
, we calculate the specific dissipation function in a generalized version of
the Becker model parameterized by $\nu \in \lbrack 0,1]$ that includes the
original Becker model for $\nu =1$ and for $\nu =0$ the well-known Maxwell
model. In Section \ref{Section: Jeffeys-Lomnitz model}, we calculate the
specific dissipation in a generalized version of the Lomnitz model
parameterized by $\alpha \in (-\infty ,1]$, which includes the so-called
Jeffreys--Lomnitz model for $\alpha \in \lbrack 0,1]$, and for $\alpha <0$
its extension. This extended model allows a transition from the limiting
case of an elastic model (without dissipation of elastic energy) at $\alpha
\rightarrow -\infty $, to the Maxwell model at $\alpha =1$, passing
continuously through the Lomnitz model at $\alpha =0$.

\section{Energy Dissipation in the Becker, Lomnitz, and Lambert Models \label%
{Section: classical models}}

\subsection{Becker Model}

The dimensionless creep function for the Becker model is (\cite{MainardiBook}, Equation (2.81)):
\begin{equation}
\psi _{B}\left( t\right) :=\mathrm{Ein}\left( \frac{t}{\tau _{0}}\right)
,\quad t\geq 0,\tau _{0}>0,\
\end{equation}%
where the subscript $B$ refers to the Becker model, and the complementary
exponential integral is defined by (\cite{Olver}, Equation (6.2.3)):%
\begin{equation}
\mathrm{Ein}\left( z\right) :=\int_{0}^{z}\frac{1-e^{-\tau }}{\tau }\,d\tau
,\quad z\in
%TCIMACRO{\U{2102} }%
%BeginExpansion
\mathbb{C}
%EndExpansion
.
\end{equation}%
For simplicity, we set the scaling factor to $\tau _{0}=1$. Thus,%
\begin{equation}
\psi _{B}^{\prime }\left( t\right) =\frac{1-e^{-t}}{t},
\end{equation}%
and, according to (\cite{Prudnikov4}, Equation (2.2.4(14))), we have%
\begin{equation}
\mathcal{L}\left[ \psi _{B}^{\prime }\left( t\right) ;s\right] =\log \left(
1+\frac{1}{s}\right) .  \label{L[psi'(t)]_Becker}
\end{equation}%
Applying (\ref{Q-1_resultado_general}), we obtain the specific dissipation:
\begin{equation}
Q_{B}^{-1}\left( \omega \right) =-\frac{\Im \left( \log \left( 1-\frac{i}{%
\omega }\right) \right) }{\Re \left( 1+\log \left( 1-\frac{i}{\omega }%
\right) \right) }.
\end{equation}%
Since
\begin{equation}
\log \left( 1-\frac{i}{\omega }\right) =\frac{1}{2}\log \left( 1+\frac{1}{%
\omega ^{2}}\right) -i\,\tan ^{-1}\left( \frac{1}{\omega }\right) ,
\end{equation}%
the specific dissipation of the Becker model is given by
\begin{equation}
Q_{B}^{-1}\left( \omega \right) =\frac{\tan ^{-1}\left( \frac{1}{\omega }%
\right) }{1+\frac{1}{2}\log \left( 1+\frac{1}{\omega ^{2}}\right) }.
\label{Q-1_B}
\end{equation}%
It is worth noting that (\ref{Q-1_B}) can also be found in
\cite{StrickMainardi}, Equation 8 and
\cite{MainardiBecker}, Equation (5.5).

Next, we derive the asymptotic behavior of $Q_{B}^{-1}\left( \omega \right)
$. On the one hand, performing the change in variable $x=\frac{1}{\omega }$
and taking into account the asymptotic formulas for $x\rightarrow 0$ (\cite{Olver}, Equations (4.6.1) and (4.24.3)):
\begin{eqnarray}
\tan ^{-1}x &\approx &x,\quad  \\
\log \left( 1+x\right)  &\approx &x,
\end{eqnarray}%
is easy to obtain%
\begin{equation}
Q_{B}^{-1}\left( \omega \right) \approx \frac{1}{\omega },\quad \omega
\rightarrow +\infty ,  \label{Q-1_Becker_w->inf}
\end{equation}%
which corresponds to the specific dissipation of the Maxwell model, see (\cite{MainardiBook}, Equation (2.79))  {taking $\tau _{0}=1$ (as we have assumed before).
On the other hand, knowing that (\cite{Olver}, Table 4.23.1) 
%MDPI: Please check whether this refers to the Table of this paper, if so, please add the table in the main text, and put it after the first citation AUTHOR -> Revision done

\begin{equation}
\lim_{x\rightarrow +\infty }\tan ^{-1}x=\frac{\pi }{2},
\end{equation}%
and taking into account that%
\begin{equation}
\log \left( 1+\frac{1}{\omega ^{2}}\right) \approx -2\log \omega ,\quad
\omega \rightarrow 0^{+},
\end{equation}%
we conclude%
\begin{equation}
Q_{B}^{-1}\left( \omega \right) \approx -\frac{\pi }{2\log \omega },\quad
\omega \rightarrow 0^{+}.  \label{Q-1_Becker_w->0}
\end{equation}

\subsection{Lomnitz Model}

The dimensionless creep function for the Lomnitz model is (\cite{MainardiBook}, Equation (2.84)):
\begin{equation}
\psi _{L}\left( t\right) :=\log \left( 1+\frac{t}{\tau _{0}}\right) ,\quad
t\geq 0,\tau _{0}>0,
\end{equation}%
where the subscript $L$ refers to the Lomnitz model. For simplicity, we set
the scaling factor to $\tau _{0}=1$. Thus,%
\begin{equation}
\psi _{L}^{\prime }\left( t\right) =\frac{1}{1+t}.
\end{equation}%
Thereby, according to (\cite{Prudnikov4}, Equation (2.1.2(5))), we have
\begin{equation}
\mathcal{L}\left[ \psi _{L}^{\prime }\left( t\right) ;s\right] =-e^{s}\,%
\mathrm{Ei}\left( -s\right) ,
\end{equation}%
where $\mathrm{Ei}\left( z\right) $ denotes the exponential integral, which
is defined by (\cite{Lebedev}, Equation (3.1.1)):
\begin{equation}
\mathrm{Ei}\left( z\right) :=\int_{-\infty }^{z}\frac{e^{t}}{t}dt,\quad
\left\vert \mathrm{arg}\,z\right\vert <\pi .
\end{equation}%
The exponential integral satisfies the following property for $x>0$ (\cite{Lebedev}, Equation (3.3.4)):%
\begin{equation}
\mathrm{Ei}\left( -i\,x\right) =\mathrm{Ci}\left( x\right) +i\left[ \frac{%
\pi }{2}-\mathrm{Si}\left( x\right) \right] ,  \label{Ei(-ix)}
\end{equation}%
where the sine and cosine integrals are defined by (\cite{Lebedev}, Equation (3.3.1))%
:
\begin{equation}
\mathrm{Si}\left( z\right) :=\int_{0}^{z}\frac{\sin t}{t}dt,\quad \mathrm{Ci}%
\left( z\right) :=\int_{\infty }^{z}\frac{\cos t}{t}dt,\quad \left\vert
\mathrm{arg}\,z\right\vert <\pi .
\end{equation}%
Hence, for $\omega >0$, we obtain
\begin{eqnarray}
\mathcal{L}\left[ \psi _{L}^{\prime }\left( t\right) ;i\,\omega \right]
&=&-e^{i\omega }\,\mathrm{Ei}\left( -i\omega \right)   \label{L[psi'_L(t)]}
\\
&=&-\left( \cos \omega +i\sin \omega \right) \left[ \mathrm{Ci}\left( \omega
\right) +i\left( \frac{\pi }{2}-\mathrm{Si}\left( \omega \right) \right) %
\right]  \\
&=&-\cos \omega \,\,\mathrm{Ci}\left( \omega \right) +\sin \omega \left(
\frac{\pi }{2}-\mathrm{Si}\left( \omega \right) \right)   \nonumber \\
&&+i\left[ -\sin \omega \,\mathrm{Ci}\left( \omega \right) -\cos \omega
\left( \frac{\pi }{2}-\mathrm{Si}\left( \omega \right) \right) \right] .
\end{eqnarray}%
Consequently, by (\ref{Q-1_resultado_general}), the specific dissipation for
the Lomnitz model is%
\begin{equation}
Q_{L}^{-1}\left( \omega \right) =\frac{\cos \omega \left( \frac{\pi }{2}-%
\mathrm{Si}\left( \omega \right) \right) +\sin \omega \,\mathrm{Ci}\left(
\omega \right) }{1-\cos \omega \,\,\mathrm{Ci}\left( \omega \right) +\sin
\omega \left( \frac{\pi }{2}-\mathrm{Si}\left( \omega \right) \right) }.
\label{Q-1_L}
\end{equation}%
{Equation (\ref{Q-1_L})\ first appeared in (\cite{LomnitzQ}, Equation (33)), although
it contains a typographical error.}
This typographical error was  corrected in \cite[Eqn. 5]{StrickMainardi}.

Next, we derive the asymptotic behavior of $Q_{L}^{-1}\left( \omega \right)
$. On the one hand, according to (\cite{Olver}, Equations (6.2.18), (6.2.19), (6.12.3) and (6.12.4)), we
obtain the following asymptotic formulas as $z\rightarrow +\infty $:
\begin{eqnarray}
\mathrm{Si}\left( z\right)  &\approx &\frac{\pi }{2}-\frac{\cos z}{z}-\frac{%
\sin z}{z^{2}}, \\
\mathrm{Ci}\left( z\right)  &\approx &\frac{\sin z}{z}-\frac{\cos z}{z^{2}}.
\end{eqnarray}%
Thus,
\begin{equation}
Q_{L}^{-1}\left( \omega \right) \approx \frac{1}{\omega },\quad \omega
\rightarrow +\infty ,  \label{Q-1_Lomnitz_w->inf}
\end{equation}%
{which corresponds to the specific dissipation of the Maxwell model, see (\cite{MainardiBook}, Equation (2.79))} taking $\tau _{0}=1$ (as we have assumed before).
On the other hand, taking into account the following power expansions valid
for $\left\vert z\right\vert <\infty $ (\cite{Olver}, Equations (6.6.5) and (6.6.6)):%
\begin{eqnarray}
\mathrm{Si}\left( z\right)  &=&\sum_{n=0}^{\infty }\frac{\left( -1\right)
^{n}\,z^{2n+1}}{\left( 2n+1\right) \,\left( 2n+1\right) !}, \\
\mathrm{Ci}\left( z\right)  &=&\gamma +\log z+\sum_{n=1}^{\infty }\frac{%
\left( -1\right) ^{n}\,z^{2n}}{2n\,\left( 2n\right) !\,},
\end{eqnarray}%
where $\gamma $ denotes the Euler--Mascheroni constant, we obtain%
\begin{equation}
Q_{L}^{-1}\left( \omega \right) \approx -\frac{\pi }{2\log \omega },\quad
\omega \rightarrow 0^{+},  \label{Q-1_Lomnitz_w->0}
\end{equation}%
which coincides with the asymptotic behavior of the Becker model obtained
in (\ref{Q-1_Becker_w->0}).

\subsection{Lambert Model}

The dimensionless creep function for the Lambert model is \cite{LambertJL}:%
\begin{equation}
\psi _{W}\left( t\right) :=W_{0}\left( t\right) ,\quad t\geq 0,
\end{equation}%
where $W_{0}\left( t\right) $ is the principal branch of the Lambert $W$
function (\cite{Olver}, Section 4.13). From (\ref{Q-1_resultado_general_2}),
the specific dissipation is given by%
\begin{equation}
Q_{W}^{-1}\left( \omega \right) =-\frac{\omega \,\mathcal{F}_{C}\left[
W_{0}\left( t\right) ;\omega \right] }{\omega \,\mathcal{F}_{S}\left[
W_{0}\left( t\right) ;\omega \right] +\sqrt{\frac{2}{\pi }}}.  \label{Q-1_W}
\end{equation}%
To the best of our knowledge, the expression in (\ref{Q-1_W})\ is new.

\subsection{Specific Dissipation in Comparison}

Figure \ref{Figure: Loss_tangent_BLW} shows the specific dissipation of the
Becker, Lomnitz, and Lambert models for $\omega \in \left( 0.01,100\right) $
on a log--log scale. Note that $Q^{-1}\left( \omega \right) $ is quite
similar for all three models, and the greatest discrepancy occurs around $%
\omega \approx 1$. This is due to the asymptotic behavior of the specific dissipation
as $\omega \rightarrow +\infty$ and $\omega \rightarrow 0^{+}$, i.e., (\ref{Q-1_Becker_w->inf}) and (\ref{Q-1_Becker_w->0}) for the Becker model,
as well as (\ref{Q-1_Lomnitz_w->inf}) and (\ref{Q-1_Lomnitz_w->0}) for the Lomnitz model.
Surprisingly, the Lambert model numerically exhibits the same asymptotic behaviors, although this is not obvious from the analytic expression given in (\ref{Q-1_W}).\vspace{-3pt}
\begin{figure}[htbp]
\includegraphics[width=0.95\textwidth]{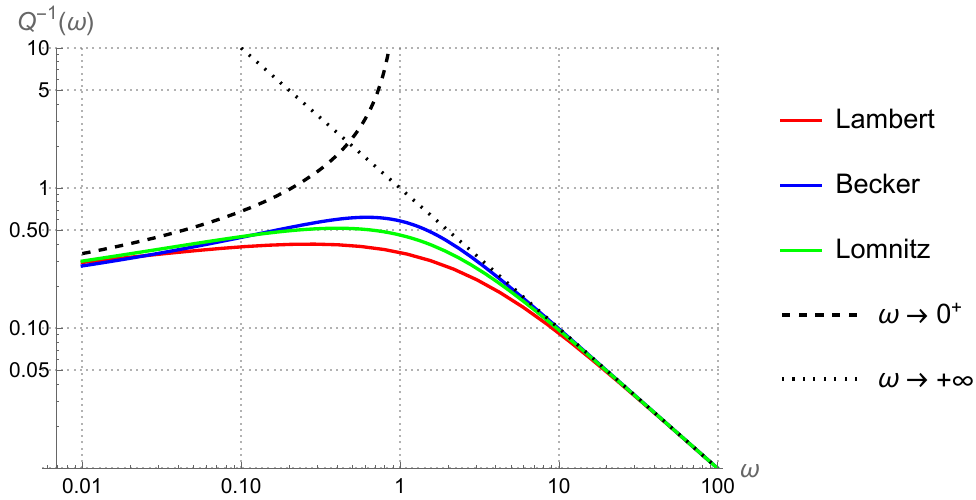}
\caption{Specific dissipation of the Becker, Lomnitz, and Lambert models.}
\label{Figure: Loss_tangent_BLW}
\end{figure}

\section{Generalized Becker Model \label{Section: Becker generalization}}

We now consider the following generalization of the Becker model by
introducing the following dimensionless creep function, parameterized by the
real parameter $\nu \in \left( 0,1\right] $ \cite{MainardiBecker}:
\begin{equation}
\psi _{\nu }\left( t\right) :=\Gamma \left( \nu +1\right) \,\mathrm{Ein}%
_{\nu }\left( \frac{t}{\tau _{0}}\right) ,\quad t\geq 0,\tau _{0}>0,
\end{equation}%
where%
\begin{equation}
\mathrm{Ein}_{\nu }\left( t\right) :=\int_{0}^{t}\frac{1-\mathrm{E}_{\nu
}\left( -\tau ^{\nu }\right) }{\tau ^{\nu }}\,d\tau ,
\end{equation}%
and $\mathrm{E}_{\alpha }\left( z\right) $ denotes the one-parameter
Mittag-Leffler function (\cite{Bateman}, Equation (18.1(1))):
\begin{equation}
\mathrm{E}_{\alpha }\left( z\right) :=\sum_{k=0}^{\infty }\frac{z^{k}}{%
\Gamma \left( \alpha k+1\right) },\quad \alpha >0,\ z\in
%TCIMACRO{\U{2102} }%
%BeginExpansion
\mathbb{C}
%EndExpansion
.
\end{equation}%
Note that, for $\nu =1$, the generalized Becker model reduces to the Becker
model
\begin{equation}
\left. \psi _{\nu }\left( t\right) \right\vert _{\nu =1}=\psi _{B}\left(
t\right) ,
\end{equation}%
since%
\begin{equation}
\mathrm{E}_{1}\left( z\right) =\sum_{k=0}^{\infty }\frac{z^{k}}{k!}=e^{z}.
\end{equation}%
As in the Becker model, we henceforth set the scaling factor to $\tau _{0}=1$.
Differentiating, we obtain
\begin{equation}
\psi _{\nu }^{\prime }\left( t\right) =\Gamma \left( \nu +1\right) \,\frac{1-%
\mathrm{E}_{\nu }\left( -t^{\nu }\right) }{t^{\nu }}=\Gamma \left( \nu
+1\right) \,\mathrm{E}_{\nu ,\nu +1}\left( -t^{\nu }\right) ,
\label{Psi'(t)_nu}
\end{equation}%
where $\mathrm{E}_{\alpha ,\beta }\left( z\right) $ denotes the
two-parameter Mittag-Leffler function, defined as
% \linebreak  \mbox{
{(\cite{Bateman}, Equation (18.1(19)):
\begin{equation}
\mathrm{E}_{\alpha ,\beta }\left( z\right) :=\sum_{k=0}^{\infty }\frac{z^{k}%
}{\Gamma \left( \alpha k+\beta \right) },\quad \alpha ,\beta >0,\ z\in
%TCIMACRO{\U{2102} }%
%BeginExpansion
\mathbb{C}
%EndExpansion
.  \label{ML_two_parameter_def}
\end{equation}%
Therefore, according to (\ref{Q-1_resultado_general}), the specific
dissipation is%
\begin{equation}
Q_{\nu }^{-1}\left( \omega \right) =-\frac{\Gamma \left( \nu +1\right) \,\Im
\left( \mathcal{L}\left[ \mathrm{E}_{\nu ,\nu +1}\left( -t^{\nu }\right)
;i\,\omega \right] \right) }{1+\Gamma \left( \nu +1\right) \,\Re \left(
\mathcal{L}\left[ \mathrm{E}_{\nu ,\nu +1}\left( -t^{\nu }\right) ;i\,\omega %
\right] \right) }.  \label{Q-1_Becker_nu}
\end{equation}%
In Appendix \ref{Appendix_A}, we define the function, for $\nu =\frac{p}{q}%
\in
%TCIMACRO{\U{211a} }%
%BeginExpansion
\mathbb{Q}
%EndExpansion
\cap \left( 0,1\right] $:%
\begin{equation}
F_{p,q}\left( s\right) :=\mathcal{L}\left[ \mathrm{E}_{p/q,1+p/q}\left(
-t^{p/q}\right) ;s\right] ,
\end{equation}%
and its computation is given in terms of the generalized hypergeometric
function defined in (\ref{pFq_def}), yielding (\ref{F_p,q_Appendix}),
\begin{eqnarray}
F_{p,q}\left( s\right)  &=&\frac{1}{p^{p/q}\,s}\sum_{j=0}^{q-1}\frac{%
\prod\limits_{\ell =0}^{p-1}\Gamma \left( \frac{\ell +1}{p}+\frac{j}{q}%
\right) }{\prod\limits_{\ell =0}^{p-1}\Gamma \left( \frac{\ell +1}{p}+\frac{%
j+1}{q}\right) }\left( -s^{-p/q}\right) ^{j}\,  \label{F_p,q_resultado} \\
&&_{p+1}F_{p}\left(
\begin{array}{c}
1,\frac{1}{p}+\frac{j}{q},\frac{2}{p}+\frac{j}{q},\ldots ,1+\frac{j}{q} \\
\frac{1}{p}+\frac{j+1}{q},\frac{2}{p}+\frac{j+1}{q},\ldots ,1+\frac{j+1}{q}%
\end{array}%
;\frac{\left( -1\right) ^{q}}{s^{p}}\right) .  \nonumber
\end{eqnarray}%
Using Mathematica, we obtain the following explicit expressions for $%
F_{p,q}\left( s\right) $:
\begin{eqnarray}
F_{1,1}\left( s\right)  &=&\log \left( 1+\frac{1}{s}\right) , \\
F_{1,2}\left( s\right)  &=&\sqrt{\frac{\pi }{s}}-\frac{2\sec ^{-1}\left(
\sqrt{s}\right) }{\sqrt{\pi \left( s-1\right) }}, \\
F_{1,3}\left( s\right)  &=&\frac{1}{s}\left\{ \frac{1}{\Gamma \left( \frac{4%
}{3}\right) }\,_{2}F_{1}\left(
\begin{array}{c}
1,1 \\
\frac{4}{3}%
\end{array}%
;\frac{-1}{s}\right) +\frac{\Gamma \left( \frac{4}{3}\right) }{\Gamma \left(
\frac{5}{3}\right) s^{1/3}}\,_{2}F_{1}\left(
\begin{array}{c}
1,\frac{4}{3} \\
\frac{5}{3}%
\end{array}%
;\frac{-1}{s}\right) \right.  \\
&&+\left. \frac{3}{2}\,\Gamma \left( \frac{5}{3}\right) s^{1/3}\left[ 1-%
\frac{s\left( 1+\frac{1}{s}\right) ^{1/3}}{s+1}\right] \right\} ,  \nonumber
\\
F_{2,3}\left( s\right)  &=&\frac{27}{112\,s^{5/3}} \\
&&\left\{ \frac{\sqrt{3}}{\pi }\left[ 3\,s^{2/3}\Gamma \left( \frac{10}{3}%
\right) \,_{3}F_{2}\left(
\begin{array}{c}
\frac{1}{2},1,1 \\
\frac{5}{6},\frac{4}{3}%
\end{array}%
;\frac{-1}{s^{2}}\right) -7\,\Gamma ^{2}\left( \frac{5}{3}\right)
\,_{3}F_{2}\left(
\begin{array}{c}
\frac{5}{6},1,\frac{4}{3} \\
\frac{7}{6},\frac{5}{3}%
\end{array}%
;\frac{-1}{s^{2}}\right) \right] \right.   \nonumber \\
&&+\left. 4\,\Gamma \left( \frac{10}{3}\right) s^{4/3}\left[ 1-\frac{\cos
\left( \frac{\cot ^{-1}s}{3}\right) ^{1/3}}{\left( 1+\frac{1}{s^{2}}\right)
^{1/6}}\right] \right\} .  \nonumber
\end{eqnarray}%
Therefore, according to (\ref{Q-1_resultado_general}), we have
\begin{equation}
\left. Q_{\nu}^{-1}\left( \omega \right)\right|_{\nu=p/q \in \mathbb{Q}} =-\frac{\Gamma \left( 1+\frac{p}{q}\right)
\,\Im \left( F_{p,q}\left( i\,\omega \right) \right) }{1+\Gamma \left( 1+%
\frac{p}{q}\right) \,\Re \left( \,F_{p,q}\left( i\,\omega \right) \right) }.
\label{Q-1_p,q}
\end{equation}

Note that the limiting case $\nu \rightarrow 0$ requires special attention,
because, according to (\ref{Psi'(t)_nu}), we obtain an undefined series:%
\begin{equation}
\lim_{\nu \rightarrow 0}\psi _{\nu }^{\prime }\left( t\right) =\lim_{\nu
\rightarrow 0}\,\mathrm{E}_{\nu ,\nu +1}\left( -t^{\nu }\right)
=\sum_{k=0}^{\infty }\left( -1\right) ^{k}=1-1+1-1+\cdots
\end{equation}%
However, by assigning the Ces\`{a}ro sum to the divergent series above,
known as Grandi's series (the %MDPI: This paper is not allowed footer format, so we move it into the main text. Please confirm it AUTHOR -> Revision confirmed
{Grandi's series is formally the value of the Dirichlet eta function (\cite{Atlas}, Equation (3:6:3))} it is %MDPI EE: Please check that the intended meaning has been retained.
 evaluated at zero, that is, $\eta\left(0\right)$.
According to (\cite{Atlas}, Section 3:7), $\eta\left(0\right) = \frac{1}{2}$. The eta function is part of a broad class of function series, known as Dirichlet series, which
have recently found new physical applications in the so-called Bessel models. For the relation between the Bessel models and the fractional Maxwell model see \cite{Giusti}), we obtain:\vspace{-3pt}
\begin{equation}
\lim_{\nu \rightarrow 0}\psi _{\nu }^{\prime }\left( t\right) =\frac{1}{2}.
\end{equation}%
Therefore, applying the Laplace transform (\ref{L[t^a;s]}), we have
\begin{equation}
\lim_{\nu \rightarrow 0}\mathcal{L}\left[ \psi _{\nu }^{\prime }\left(
t\right) ;i\,\omega \right] =-\frac{i}{2\omega }.
\end{equation}%
Thus, from (\ref{Q-1_resultado_general}), the specific dissipation is
\begin{equation}
\lim_{\nu \rightarrow 0}Q_{\nu }^{-1}\left( \omega \right) =-\lim_{\nu
\rightarrow 0}\frac{\Im \left( \mathcal{L}\left[ \psi _{\nu }^{\prime
}\left( t\right) ;i\,\omega \right] \right) }{1+\Re \left( \mathcal{L}\left[
\psi _{\nu }^{\prime }\left( t\right) ;i\,\omega \right] \right) }=\frac{1}{%
2\omega }.  \label{Q-1_0_resultado}
\end{equation}%
{It is noteworthy that (\ref{Q-1_0_resultado}) corresponds to the Maxwell
model and appears in (\cite{MainardiBecker}, Equation (5.4))}  taking $q=1$ (as we have assumed from the beginning).

\begin{figure}[htbp]
\includegraphics[width=0.95\textwidth]{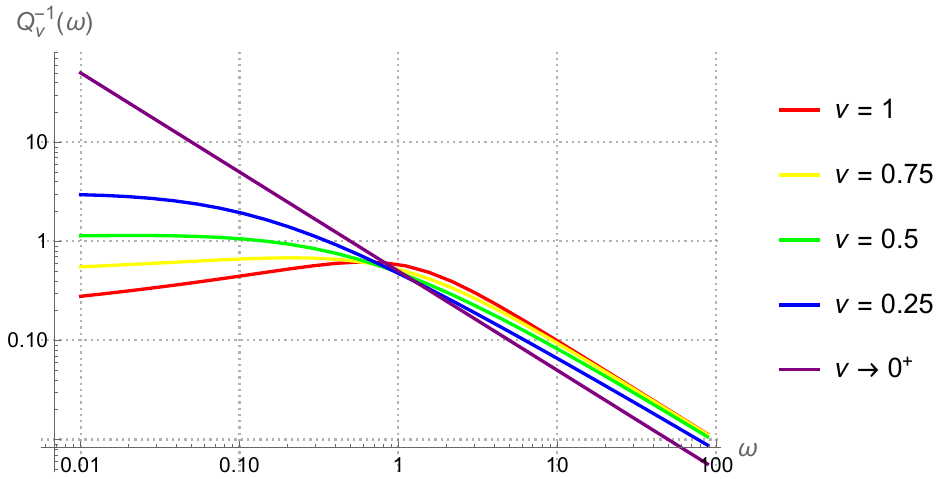}
\caption{Specific  dissipation of the generalized Becker model.}
\label{Figure: Fractional Becker}
\end{figure}

Figure \ref{Figure: Fractional Becker} shows the specific dissipation of the
generalized Becker model $Q_{\nu }^{-1}\left( \omega \right) $ for different
values of $\nu \in \left( 0,1\right] $, and for $\omega \in \left(
0.01,100\right) $ on a log-log scale. Note that the greatest discrepancy
among the curves corresponding to different values of $\nu $ is observed for
small values of the frequency, i.e., $0<\omega <1$.\vspace{-3pt}

\section{The Extended Jeffreys--Lomnitz Model \label{Section: Jeffeys-Lomnitz
model}}

According to (\cite{MainardiBook}, Equation (2.94)), the Lomnitz model can be
extended as follows for $\alpha \in \left( -\infty ,1\right] $:\
\begin{equation}
\psi _{\alpha }\left( t\right) :=\frac{1}{\alpha }\left[ \left( 1+\frac{t}{%
\tau _{0}}\right) ^{\alpha }-1\right] ,\quad t\geq 0,\tau _{0}>0.
\end{equation}%
Note that the intermediate case $\alpha \rightarrow 0$ corresponds to the
Lomnitz model:%
\begin{equation}
\lim_{\alpha \rightarrow 0}\psi _{\alpha }\left( t\right) =\log \left( 1+%
\frac{t}{\tau _{0}}\right) =\psi _{L}\left( t\right) .  \label{alpha->0}
\end{equation}%
Again, for simplicity, we set the scaling factor to $\tau _{0}=1$. Thus,%
\begin{equation}
\psi _{\alpha }^{\prime }\left( t\right) =\left( 1+t\right) ^{\alpha -1}.
\label{Psi'_a(t)}
\end{equation}%
Using the definition of the Laplace transform (\ref{Laplace_def}) and
performing the substitution $\tau =1+t$, we obtain
\begin{eqnarray}
\mathcal{L}\left[ \psi _{\alpha }^{\prime }\left( t\right) ;i\omega \right]
&=&\int_{0}^{\infty }\frac{e^{-i\omega t}}{\left( 1+t\right) ^{1-\alpha }}dt
\\
&=&\underset{\Re \left( \mathcal{L}\left[ \psi _{\alpha }^{\prime }\left(
t\right) ;i\omega \right] \right) }{\underbrace{\int_{0}^{\infty }\frac{\cos
\omega t}{\left( 1+t\right) ^{1-\alpha }}dt}}-i\underset{-\Im \left(
\mathcal{L}\left[ \psi _{\alpha }^{\prime }\left( t\right) ;i\omega \right]
\right) }{\underbrace{\int_{0}^{\infty }\frac{\sin \omega t}{\left(
1+t\right) ^{1-\alpha }}dt}}.  \nonumber
\end{eqnarray}%
According to (\ref{Q-1_resultado_general}), the specific dissipation is%
\begin{equation}
Q_{\alpha }^{-1}\left( \omega \right) =-\frac{\Im \left( \mathcal{L}\left[
\psi _{\alpha }^{\prime }\left( t\right) ;i\,\omega \right] \right) }{1+\Re
\left( \mathcal{L}\left[ \psi _{\alpha }^{\prime }\left( t\right) ;i\,\omega %
\right] \right) }.  \label{Q-1_a}
\end{equation}%
Applying the results in Appendix \ref{Appendix_B}, i.e., (\ref{IA_resultado})
and (\ref{IB_resultado}), we obtain for $\alpha \in \left( 0,1\right] $,\vspace{-3pt}
\begin{eqnarray}
&&Q_{\alpha }^{-1}\left( \omega \right)  \label{Q-1_a_Hyper} \\
&=&\frac{\Gamma \left( \alpha +2\right) \sin \left( \frac{\alpha \pi }{2}%
-\omega \right) +\omega ^{\alpha +1}\,_{1}F_{2}\left(
\begin{array}{c}
1 \\
1+\frac{\alpha }{2},\frac{\alpha +3}{2}%
\end{array}%
;-\frac{\omega ^{2}}{4}\right) }{\alpha \left( \alpha +1\right) \,\omega
^{\alpha }+\Gamma \left( \alpha +2\right) \cos \left( \frac{\alpha \pi }{2}%
-\omega \right) -\left( \alpha +1\right) \,\omega ^{\alpha
}\,_{1}F_{2}\left(
\begin{array}{c}
1 \\
\frac{\alpha +1}{2},1+\frac{\alpha }{2}%
\end{array}%
;-\frac{\omega ^{2}}{4}\right) }.  \nonumber
\end{eqnarray}

As an alternative representation for $Q_{\alpha }^{-1}\left( \omega \right) $%
, we can compute the Laplace transform of (\ref{Psi'_a(t)}) using (\cite{Prudnikov4}, Equation
(2.1.2(1))):%
\begin{equation}
\mathcal{L}\left[ \psi _{\alpha }^{\prime }\left( t\right) ;s\right] =%
\mathcal{L}\left[ \left( 1+t\right) ^{\alpha -1};s\right] =e^{s}\,s^{-\alpha
}\,\Gamma \left( \alpha ,s\right) ,
\end{equation}%
where
\begin{equation}
\Gamma \left( a,z\right):= \int_{z}^{\infty}t^{a-1}e^{-t}\,dt, \quad z\in \mathbb{C},
\end{equation}
denotes the upper incomplete gamma function (\cite{Olver}, Equation (8.2.2)). Therefore, from (\ref%
{Q-1_resultado_general}), for $\alpha \in \left( -\infty ,1\right] $, the
specific dissipation is
\begin{equation}
Q_{\alpha }^{-1}\left( \omega \right) =-\frac{\Im \left( e^{i\omega
}\,\left( i\,\omega \right) ^{-\alpha }\,\Gamma \left( \alpha ,i\,\omega
\right) \right) }{1+\Re \left( e^{i\omega }\,\left( i\,\omega \right)
^{-\alpha }\,\Gamma \left( \alpha ,i\,\omega \right) \right) }.
\label{Q-1_a_Gamma}
\end{equation}
According to our numerical experiments, both expressions (\ref{Q-1_a_Hyper}) and (\ref{Q-1_a_Gamma})
are numerically equivalent for $\alpha \in \left( 0,1\right] $,
but (\ref{Q-1_a_Gamma}) is significantly faster to evaluate than (\ref{Q-1_a_Hyper});
see the Mathematica notebook available at: %
\url{https://shorturl.at/KUDgC} accessed on 11 August 2026. %MDPI: 1. Please add the access date (format: Date Month Year), e.g., accessed on 1 January 2020. 2. Please manually check if the following URL is truly invalid, if so, please revise it. AUTHOR -> Revision done

As noted in (\ref{alpha->0}), the limiting case $\alpha \rightarrow 0$
corresponds to the Lomnitz model. Indeed, applying the following property
valid for $x>0$ (\cite{Atlas}, Equation (45:9:7)):%
\begin{equation}
\lim_{\nu \rightarrow 0}\Gamma \left( \nu ,x\right) =-\mathrm{Ei}\left(
-x\right) ,
\end{equation}%
to (\ref{Q-1_a_Gamma}), we obtain
\begin{equation}
\lim_{\alpha \rightarrow 0}Q_{\alpha }^{-1}\left( \omega \right) =\frac{\Im
\left( e^{i\omega }\,\,\mathrm{Ei}\left( -i\omega \right) \right) }{1-\Re
\left( e^{i\omega }\,\,\mathrm{Ei}\left( -i\omega \right) \right) }.
\end{equation}%
Recalling that $\omega >0$, and taking into account (\ref{L[psi'_L(t)]}), we
arrive at (\ref{Q-1_L}), i.e.,
\begin{equation}
\lim_{\alpha \rightarrow 0}Q_{\alpha }^{-1}\left( \omega \right)
=Q_{L}^{-1}\left( \omega \right) .  \label{Limit_Q-1_a->0}
\end{equation}
However, this consistency check is highly non-trivial from the expression
given in (\ref{Q-1_a_Hyper}). Both expressions, (\ref{Q-1_a_Hyper})\ and (%
\ref{Q-1_a_Gamma}), are computationally efficient. However, (\ref%
{Q-1_a_Hyper})\ is valid only for $\alpha \in \left( 0,1\right] $, whereas (%
\ref{Q-1_a_Gamma}) remains valid for negative values of $\alpha $.

The case $\alpha =1$ deserves special attention, because from (\ref%
{Psi'_a(t)}) we have
\begin{equation}
\left. \psi _{\alpha }^{\prime }\left( t\right) \right\vert _{\alpha =1}=1.
\end{equation}%
Therefore, from (\ref{Q-1_a}) and applying the Laplace transform (\ref%
{L[t^a;s]}), the specific dissipation reduces to
\begin{equation}
\left. Q_{\alpha }^{-1}\left( \omega \right) \right\vert _{\alpha =1}=\frac{1%
}{\omega },  \label{Q-1_a=1_Maxwell}
\end{equation}
{which corresponds to the specific dissipation of the Maxwell model, see (\cite{MainardiBook},
Equation (2.79))} taking $\tau_{0} = 1$ (as we have assumed before). As a consistency test, it is worth noting that (%
\ref{Q-1_a=1_Maxwell}) can be derived from (\ref{Q-1_a_Hyper}), applying
the reduction formulas (\cite{Prudnikov3}, Equation (7.14.2(78)) and (7.13.1(6))):\vspace{-3pt}
\begin{eqnarray}
_{1}F_{2}\left(
\begin{array}{c}
1 \\
\frac{3}{2},2%
\end{array}%
;z^{2}\right) &=&\frac{\cosh \left( 2z\right) -1}{2z^{2}}, \\
_{0}F_{1}\left(
\begin{array}{c}
\text{---} \\
\frac{3}{2}%
\end{array}%
;-z\right) &=&\frac{\sin \left( 2\sqrt{z}\right) }{2\sqrt{z}}.
\end{eqnarray}

Note also that the asymptotic behavior of $Q_{\alpha }^{-1}\left( \omega
\right) $ as $\omega \rightarrow +\infty $ corresponds to the 
{Maxwell model (%
\ref{Q-1_a=1_Maxwell}). Indeed, using the asymptotic formula (\cite{Atlas}, Equation (45:9:5))}:%
\begin{equation}
\Gamma \left( a,z\right) \approx \frac{e^{-z}}{z^{1-a}},\quad z\rightarrow
\infty , \quad \left\vert \mathrm{arg}\,z\right\vert <\frac{3\pi}{2},
\end{equation}%
from (\ref{Q-1_a_Gamma}), we conclude%
\begin{equation}
Q_{\alpha }^{-1}\left( \omega \right) \approx \frac{1}{\omega },\quad \omega
\rightarrow +\infty .  \label{Q-1_a_Asymp}
\end{equation}
Note that (\ref{Q-1_a_Asymp}) generalizes the asymptotic behavior obtained in (\ref{Q-1_Lomnitz_w->inf}) for the Lomnitz model.

Figure \ref{Figure: Loss tangent extended Lomnitz} shows the specific
dissipation of the extended Jeffreys--Lomnitz model $Q_{\alpha }^{-1}\left(
\omega \right) $ for different values of $\alpha \in \left( -\infty ,1\right]
$, and for $\omega \in \left( 0.01,100\right) $ on a log--log scale. Note
that the greatest discrepancy among the curves corresponding to different
values of $\alpha $ is observed for small values of the frequency, i.e., $%
0<\omega <1$. Notice as well that Figure \ref{Figure: Loss tangent extended
Lomnitz} confirms that the asymptotic behavior of $Q_{\alpha }^{-1}\left(
\omega \right) $ as $\omega \rightarrow +\infty $ given by (\ref{Q-1_a_Asymp}%
)\ coincides with that of the Maxwell model (\ref{Q-1_a=1_Maxwell}).
\begin{figure}[htbp]
\includegraphics[width=0.95\textwidth]{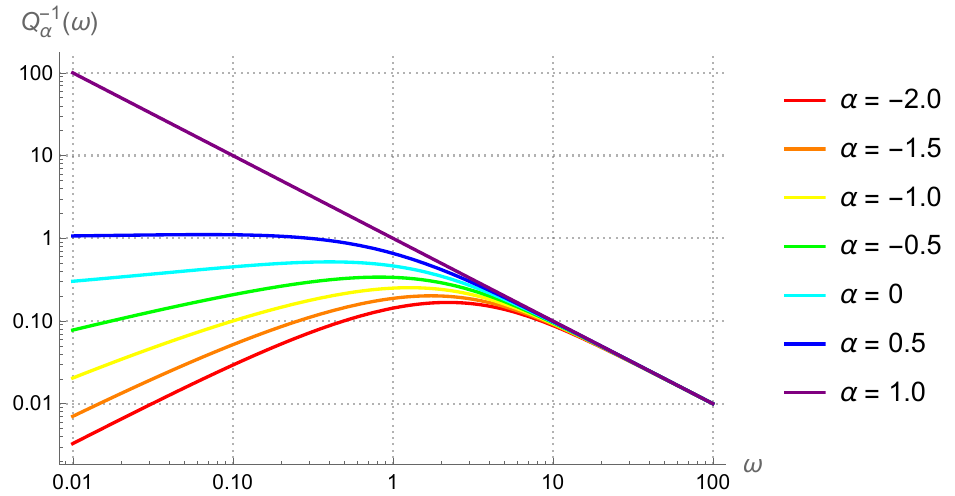}
\caption{Specific dissipation of the extended Jeffreys--Lomnitz model.}
\label{Figure: Loss tangent extended Lomnitz}
\end{figure}

\newpage
\section{Conclusions}

On the one hand, we derived general formulas for the specific dissipation $%
Q^{-1}\left( \omega \right) $ in (\ref{Q-1_resultado_general}) and (\ref%
{Q-1_resultado_general_2}) in order to calculate it for different models in
Earth rheology. These formulas simplify the calculations found in the
existing literature for the classical models, namely the Becker
and Lomnitz models, as well as yielding a simple new expression given in (\ref{Q-1_W}%
) for the recent model based on the Lambert $W$ function.
Therefore, the present study complements the comparison made in
\cite{BeckerLomnitzLambert} between these models.
Figure \ref{Figure: Loss_tangent_BLW} shows that the specific dissipation is quite
similar for the Becker, Lomnitz, and Lambert models, with the greatest discrepancy occurring around $%
\omega \approx 1$. It is worth noting that the Becker and Lomnitz models behave asymptotically
like the Maxwell model as $\omega \rightarrow +\infty$, see
(\ref{Q-1_Becker_w->inf}) and (\ref{Q-1_Lomnitz_w->inf}).
Furthermore, these models exhibit the same asymptotic behavior as
$\omega \rightarrow 0^{+}$, \mbox{see (\ref{Q-1_Becker_w->0}) and
(\ref{Q-1_Lomnitz_w->0}).}

On the other hand, from the general result for the specific dissipation given
in (\ref{Q-1_resultado_general}), we obtained novel expressions for $Q_{\nu
}^{-1}\left( \omega \right) $ for the generalized Becker model in (\ref%
{Q-1_p,q}) and (\ref{F_p,q_resultado}) when $\nu$ is rational.
The latter is not a significant limitation for visualizing how the specific dissipation varies with respect to $\nu$,
as shown in Figure \ref{Figure: Fractional Becker}.
We also obtained novel expressions for $Q_{\alpha }^{-1}\left( \omega \right) $ for
the extended Jeffreys--Lomnitz model in (\ref{Q-1_a_Hyper}) and (\ref%
{Q-1_a_Gamma}). It is worth noting that (\ref{Q-1_a_Hyper}) is valid only
for $\alpha \in \left( 0,1\right] $, whereas (\ref{Q-1_a_Gamma}) is valid
for $\alpha \in \left( -\infty ,1\right] $. Furthermore, the expressions for $%
Q_{\nu }^{-1}\left( \omega \right) $ and $Q_{\alpha }^{-1}\left( \omega
\right) $ allow us to readily plot the specific dissipation for different
values of the parameters $\nu$ and $\alpha$, respectively. \mbox{Figures \ref
{Figure: Fractional Becker} and \ref{Figure: Loss tangent extended Lomnitz}}
show that the greatest discrepancies in the specific dissipations
$Q_{\nu }^{-1}\left( \omega \right)$ and
$Q_{\alpha }^{-1}\left( \omega \right)$, respectively, correspond
to small values of the frequency (i.e., $0<\omega <1$). Moreover, Figure \ref%
{Figure: Loss tangent extended Lomnitz} confirms that the asymptotic
behavior of $Q_{\alpha }^{-1}\left( \omega \right) $ as $\omega \rightarrow
+\infty $ coincides with that of the Maxwell model.

It is important to point out that although the generalized Becker model and
the extended Jeffreys--Lomnitz model have already been proposed in the
literature, closed-form expressions for the specific dissipation have not
previously been derived.

Finally, all the graphs presented in this paper were plotted using
Mathematica 13.2. The corresponding Mathematica notebook is available at:
\url{https://shorturl.at/KUDgC} accessed on 11 August 2026. %MDPI: 1. Please add the access date (format: Date Month Year), e.g., accessed on 1 January 2020. 2. Please manually check if the following URL is truly invalid, if so, please revise it. AUTHOR -> Revision done

\section*{Authors contributions}
{Conceptualization %MDPI: Please check this section. Are all informations correct? If yes, please confirm. AUTHOR -> Revision confirmed
, F.M.; methodology, F.M.; software, J.L.G.-S.; validation, J.L.G.-S.; formal analysis, J.L.G.-S.; writing---original draft preparation, J.L.G.-S.; writing---review and editing, F.M.; supervision, F.M.; project administration, F.M. All authors have read and agreed to the published version of the manuscript.}

\section*{Funding}
{This  research received no external funding}

\section*{Acknowledgments}
The research activity of F. Mainardi  has been carried out in the framework of the activities of the National Group of Mathematical Physics (GNFM, INdAM).
\\
The authors are grateful to the anonymous referees for valuable suggestions that helped us to improve the presentation of the results.

%%%%%%%%%%%%%%%%%%%%%%%%%%%
\appendix
\section[\appendixname~\thesection]{A Useful Laplace Transform}
%\subsection[\appendixname~\thesubsection]{A Useful Laplace Transform}

%\renewcommand{\theequation}{\thesection.\arabic{equation}} %
%\setcounter{equation}{0} \numberwithin{equation}{section}

%\section[\appendixname~\thesection]{A Useful Laplace Transform}

\label{Appendix_A}

Using %(\ref{Q-1_Becker_nu}) and
(\ref{ML_two_parameter_def}), we compute:\vspace{-3pt}
\begin{equation}
\mathcal{L}\left[ \mathrm{E}_{\nu ,\nu +1}\left( -t^{\nu }\right) ;s\right]
=\sum_{k=0}^{\infty }\frac{\left( -1\right) ^{k}}{\Gamma \left( 1+\left(
k+1\right) \nu \right) }\mathcal{L}\left[ t^{\nu k};s\right] .
\end{equation}%
Applying the Laplace transform (\cite{Prudnikov4}, Equation (2.1.1(1))):%
\begin{equation}
\mathcal{L}\left[ t^{\alpha };s\right] =\frac{\Gamma \left( 1+\alpha \right)
}{s^{1+\alpha }},  \label{L[t^a;s]}
\end{equation}%
we obtain%
\begin{equation}
\mathcal{L}\left[ \mathrm{E}_{\nu ,\nu +1}\left( -t^{\nu }\right) ;s\right] =%
\frac{1}{s}\sum_{k=0}^{\infty }\frac{\Gamma \left( 1+k\nu \right) }{\Gamma
\left( 1+\left( k+1\right) \nu \right) }\left( -\frac{1}{s^{\nu }}\right)
^{k}.
\end{equation}%
Now, for $\nu =\frac{p}{q}\in
%TCIMACRO{\U{211a} }%
%BeginExpansion
\mathbb{Q}
%EndExpansion
$, define\vspace{-3pt}
\begin{equation}
F_{p,q}\left( s\right) :=\mathcal{L}\left[ \mathrm{E}_{p/q,1+p/q}\left(
-t^{p/q}\right) ;s\right] .
\end{equation}%
Thus, setting $z=-s^{-p/q}$, we have\vspace{-3pt}
\begin{eqnarray}
s\,F_{p,q}\left( s\right)  &=&\sum_{k=0}^{\infty }\frac{\Gamma \left( 1+k%
\frac{p}{q}\right) }{\Gamma \left( 1+\left( k+1\right) \frac{p}{q}\right) }%
z^{k} \\
&=&\sum_{j=0}^{q-1}\sum_{k=0}^{\infty }\frac{\Gamma \left( 1+\left(
qk+j\right) \frac{p}{q}\right) }{\Gamma \left( 1+\left( qk+j+1\right) \frac{p%
}{q}\right) }z^{qk+j}.  \nonumber
\end{eqnarray}%
Applying the multiplication formula (\cite{Olver}, Equation (5.5.6)):%
\begin{equation}
\Gamma \left( n\,z\right) =\left( 2\pi \right) ^{\left( 1-n\right)
/2}n^{n\,z-1/2}\prod\limits_{\ell =0}^{n-1}\Gamma \left( z+\frac{\ell }{n}%
\right) ,
\end{equation}%
together with definition of the Pochhammer symbol (\cite{Olver}, Equations (5.2.4) and (5.2.5)) %
\begin{equation}
\left( x\right) _{k}:=\left\{
\begin{array}{ll}
x\left( x+1\right) \left( x+2\right) \cdots \left( x+k-1\right) , & \left(
x\right) _{0}=1, \\
%TCIMACRO{\TeXButton{TeX field}{\displaystyle}}%
%BeginExpansion
\displaystyle%
%EndExpansion
\frac{\Gamma \left( x+k\right) }{\Gamma \left( x\right) },\quad  & x\neq
0,-1,-2,\ldots
\end{array}%
\right.   \label{pochhammer_def}
\end{equation}%
we obtain%
\begin{eqnarray}
s\,F_{p,q}\left( s\right)
&=&p^{-p/q}\sum_{j=0}^{q-1}z^{j}\sum_{k=0}^{\infty }\frac{\prod\limits_{\ell
=0}^{p-1}\Gamma \left( \frac{\ell +1}{p}+k+\frac{j}{q}\right) }{%
\prod\limits_{\ell =0}^{p-1}\Gamma \left( \frac{\ell +1}{p}+k+\frac{j+1}{q}%
\right) }\left( z^{q}\right) ^{k} \\
&=&p^{-p/q}\sum_{j=0}^{q-1}\frac{\prod\limits_{\ell =0}^{p-1}\Gamma \left(
\frac{\ell +1}{p}+\frac{j}{q}\right) }{\prod\limits_{\ell =0}^{p-1}\Gamma
\left( \frac{\ell +1}{p}+\frac{j+1}{q}\right) }z^{j}\sum_{k=0}^{\infty }%
\frac{\prod\limits_{\ell =0}^{p-1}\left( \frac{\ell +1}{p}+\frac{j}{q}%
\right) _{k}}{\prod\limits_{\ell =0}^{p-1}\left( \frac{\ell +1}{p}+\frac{j+1%
}{q}\right) _{k}}\left( z^{q}\right) ^{k}.
\end{eqnarray}%
{According to the definition of the generalized hypergeometric function (\cite{Olver}, Equation (16.2.1)):}
\begin{equation}
_{p}F_{q}\left(
\begin{array}{c}
a_{1},\ldots ,a_{p} \\
b_{1},\ldots ,b_{q}%
\end{array}%
;z\right) :=\sum_{k=0}^{\infty }\frac{\left( a_{1}\right) _{k}\cdots \left(
a_{p}\right) _{k}}{\left( b_{1}\right) _{k}\cdots \left( b_{q}\right) _{k}}%
\frac{z^{k}}{k!}, \quad z\in \mathbb{C}, \label{pFq_def}
\end{equation}%
the inner sum can be recast as a hypergeometric series, yielding%
\begin{eqnarray}
F_{p,q}\left( s\right)  &=&\frac{1}{p^{p/q}\,s}\sum_{j=0}^{q-1}\frac{%
\prod\limits_{\ell =0}^{p-1}\Gamma \left( \frac{\ell +1}{p}+\frac{j}{q}%
\right) }{\prod\limits_{\ell =0}^{p-1}\Gamma \left( \frac{\ell +1}{p}+\frac{%
j+1}{q}\right) }\left( -s^{-p/q}\right) ^{j}\,  \label{F_p,q_Appendix} \\
&&_{p+1}F_{p}\left(
\begin{array}{c}
1,\frac{1}{p}+\frac{j}{q},\frac{2}{p}+\frac{j}{q},\ldots ,1+\frac{j}{q} \\
\frac{1}{p}+\frac{j+1}{q},\frac{2}{p}+\frac{j+1}{q},\ldots ,1+\frac{j+1}{q}%
\end{array}%
;\frac{\left( -1\right) ^{q}}{s^{p}}\right) .  \nonumber
\end{eqnarray}

\section[\appendixname~\thesection]{Auxiliary Integrals}

\label{Appendix_B}

We want to calculate the integrals:\
\begin{equation}
I_{A}=\int_{0}^{\infty }\frac{\cos \omega t}{\left( 1+t\right) ^{1-\alpha }}%
\,dt,
\end{equation}
and%
\begin{equation}
I_{B}=\int_{0}^{\infty }\frac{\sin \omega t}{\left( 1+t\right) ^{1-\alpha }}%
\,dt.
\end{equation}

\subsection[\appendixname~\thesubsection]{First Integral}

For the first integral, perform the change in variable $\tau =1+t$ to obtain%
\begin{eqnarray}
I_{A} &=&\int_{1}^{\infty }\frac{\cos \left( \omega \tau -\omega \right) }{%
\tau ^{1-\alpha }}d\tau  \nonumber \\
&=&\underset{A_{\infty }}{\underbrace{\int_{0}^{\infty }\frac{\cos \left(
\omega \tau -\omega \right) }{\tau ^{1-\alpha }}d\tau }}-\underset{A_{1}}{%
\underbrace{\int_{0}^{1}\frac{\cos \left( \omega \tau -\omega \right) }{\tau
^{1-\alpha }}d\tau }}.  \label{Ia=A_inf+A_1}
\end{eqnarray}%
On the one hand,
\begin{equation}
A_{\infty }=\cos \omega \int_{0}^{\infty }\frac{\cos \omega \tau }{\tau
^{1-\alpha }}d\tau +\sin \omega \int_{0}^{\infty }\frac{\sin \omega \tau }{%
\tau ^{1-\alpha }}d\tau .  \label{A_inf_a}
\end{equation}%
Taking into account the definitions of the Fourier sine and cosine
transforms (\ref{Fourier_sine_def}) and (\ref{Fourier_cosine_def}), we have
\begin{eqnarray}
\int_{0}^{\infty }\frac{\cos \omega \tau }{\tau ^{1-\alpha }}d\tau &=&\sqrt{%
\frac{\pi }{2}}\,\mathcal{F}_{C}\left[ \frac{1}{t^{1-\alpha }};\omega \right]
, \\
\int_{0}^{\infty }\frac{\sin \omega \tau }{\tau ^{1-\alpha }}d\tau &=&\sqrt{%
\frac{\pi }{2}}\,\mathcal{F}_{S}\left[ \frac{1}{t^{1-\alpha }};\omega \right]
.
\end{eqnarray}%
However, according to (\cite{Gradshteyn}, Equations (17.34(1)) and (17.32(2))), we find%
\begin{eqnarray}
\mathcal{F}_{C}\left[ \frac{1}{t^{\nu }};\omega \right] &=&\sqrt{\frac{\pi }{%
2}}\frac{\omega ^{\nu -1}}{\Gamma \left( \nu \right) }\sec \left( \frac{\pi
\nu }{2}\right) ,\quad 0<\Re \left( \nu \right) <1, \\
\mathcal{F}_{S}\left[ \frac{1}{t^{\nu }};\omega \right] &=&\sqrt{\frac{2}{%
\pi }}\omega ^{\nu -1}\Gamma \left( 1-\nu \right) \cos \left( \frac{\pi \nu
}{2}\right) ,\quad 0<\Re \left( \nu \right) <2.
\end{eqnarray}%
Thus,
\begin{equation}
\int_{0}^{\infty }\frac{\cos \omega \tau }{\tau ^{1-\alpha }}d\tau =\frac{%
\pi }{2\,\omega ^{\alpha }\,\Gamma \left( 1-\alpha \right) }\csc \left(
\frac{\pi \alpha }{2}\right) .
\end{equation}%
Applying the reflection formula (\cite{Lebedev}, Equation (1.2.1)):%
\begin{equation}
\Gamma \left( z\right) \,\Gamma \left( 1-z\right) =\pi \csc \left( \pi
z\right) ,\quad z\notin
%TCIMACRO{\U{2124} }%
%BeginExpansion
\mathbb{Z}
%EndExpansion
,
\end{equation}%
we obtain%
\begin{equation}
\int_{0}^{\infty }\frac{\cos \omega \tau }{\tau ^{1-\alpha }}d\tau =\frac{%
\Gamma \left( \alpha \right) }{\omega ^{\alpha }}\cos \left( \frac{\pi
\alpha }{2}\right) .  \label{Int_cos_resultado}
\end{equation}%
Similarly,
\begin{equation}
\int_{0}^{\infty }\frac{\sin \omega \tau }{\tau ^{1-\alpha }}d\tau =\frac{%
\Gamma \left( \alpha \right) }{\omega ^{\alpha }}\sin \left( \frac{\pi
\alpha }{2}\right) ,  \label{Int_sin_resultado}
\end{equation}%
hence,
\begin{equation}
A_{\infty }=\frac{\Gamma \left( \alpha \right) }{\omega ^{\alpha}}\cos
\left( \frac{\alpha \pi }{2}-\omega \right) .  \label{A_inf_resultado}
\end{equation}%
{On the other hand, expanding the cosine function in its Maclaurin series (\cite{Olver}, Equation (4.19.2):}%
\begin{equation}
\cos z=\sum_{k=0}^{\infty }\frac{\left( -1\right) ^{k}z^{2k}}{\left(
2k\right) !},
\end{equation}%
and taking into account the definition of the beta function (\cite{Olver}, Equation (5.12.1)):%
\begin{equation}
\mathrm{B}\left( a,b\right) :=\int_{0}^{1}x^{a-1}\left( 1-x\right) ^{b-1}dx=%
\frac{\Gamma \left( a\right) \,\Gamma \left( b\right) }{\Gamma \left(
a+b\right) }, \quad \Re\left(a\right) >0,\,\, \Re\left(b\right) >0, \label{Beta_def}
\end{equation}%
we obtain
\begin{equation}
A_{1} =\sum_{k=0}^{\infty }\frac{\left( -1\right) ^{k}\omega ^{2k}}{\left(
2k\right) !}\int_{0}^{1}\tau ^{\alpha -1}\left( 1-\tau \right) ^{2k}d\tau
=\sum_{k=0}^{\infty }\frac{\left( -1\right) ^{k}\omega ^{2k}\,\Gamma \left(
\alpha \right) }{\Gamma \left( \alpha +2k+1\right) }.
\end{equation}%
According to the definition of the Pochhammer symbol (\ref{pochhammer_def}),
and taking into account the following formula (\cite{Atlas}, Equation (18:5:5)):%
\begin{equation}
\left( x\right) _{2n+1}=4^{n}x\,\left( \frac{1+x}{2}\right) _{n}\left( 1+%
\frac{x}{2}\right) _{n},
\end{equation}%
we have%
\begin{equation}
\frac{\Gamma \left( \alpha +2k+1\right) \,}{\Gamma \left( \alpha \right) }%
=\left( \alpha \right) _{2k+1}=\alpha \,4^{k}\left( \frac{\alpha +1}{2}%
\right) _{k}\left( 1+\frac{\alpha }{2}\right) _{k},
\end{equation}%
hence, we obtain%
\begin{equation}
A_{1}=\frac{1}{\alpha }\sum_{k=0}^{\infty }\frac{1}{\left( \frac{\alpha +1}{2%
}\right) _{k}\left( 1+\frac{\alpha }{2}\right) _{k}}\left( -\frac{\omega ^{2}%
}{4}\right) ^{k}.
\end{equation}%
Finally, using the definition of the generalized hypergeometric function (\ref{pFq_def}), we obtain
%is defined by
%\cite[Eqn. 16.2.1]{Olver}:
%\begin{equation}
%_{p}F_{q}\left(
%\begin{array}{c}
%a_{1},\ldots ,a_{p} \\
%b_{1},\ldots ,b_{q}%
%\end{array}%
%;z\right) =\sum_{k=0}^{\infty }\frac{\left( a_{1}\right) _{k}\cdots \left(
%a_{p}\right) _{k}}{\left( b_{1}\right) _{k}\cdots \left( b_{q}\right) _{k}}%
%\frac{z^{k}}{k!},  \label{Hypergeometric_def}
%\end{equation}%
%we obtain
\begin{equation}
A_{1}=\frac{1}{\alpha }\,_{1}F_{2}\left(
\begin{array}{c}
1 \\
\frac{\alpha +1}{2},1+\frac{\alpha }{2}%
\end{array}%
;-\frac{\omega ^{2}}{4}\right) .  \label{A_1_resultado}
\end{equation}%
Substituting the results (\ref{A_1_resultado})\ and (\ref{A_inf_resultado})\
into (\ref{Ia=A_inf+A_1}), we obtain
\begin{equation}
\int_{0}^{\infty }\frac{\cos \omega t}{\left( 1+t\right) ^{1-\alpha }}\,dt =%
\frac{\Gamma \left( \alpha \right) }{\omega ^{\alpha}}\cos \left( \frac{%
\alpha \pi }{2}-\omega \right) -\frac{1}{\alpha }\,_{1}F_{2}\left(
\begin{array}{c}
1 \\
\frac{\alpha +1}{2},1+\frac{\alpha }{2}%
\end{array}%
;-\frac{\omega ^{2}}{4}\right) .  \label{IA_resultado}
\end{equation}

\subsection[\appendixname~\thesubsection]{Second Integral}

For the second integral, perform the change in variable $\tau =1+t$ to obtain%
\begin{eqnarray}
I_{B} &=&\int_{1}^{\infty }\frac{\sin \left( \omega \tau -\omega \right) }{%
\tau ^{1-\alpha }}d\tau  \nonumber \\
&=&\underset{B_{\infty }}{\underbrace{\int_{0}^{\infty }\frac{\sin \left(
\omega \tau -\omega \right) }{\tau ^{1-\alpha }}d\tau }}-\underset{B_{1}}{%
\underbrace{\int_{0}^{1}\frac{\sin \left( \omega \tau -\omega \right) }{\tau
^{1-\alpha }}d\tau }}.  \label{Ib=B_1+B_inf}
\end{eqnarray}%
On the one hand, taking into account the results given in (\ref%
{Int_cos_resultado})\ and (\ref{Int_sin_resultado}), we have
\begin{equation}
B_{\infty } =\cos \omega \int_{0}^{\infty }\frac{\sin \omega \tau }{\tau
^{1-\alpha }}d\tau -\sin \omega \int_{0}^{\infty }\frac{\cos \omega \tau }{%
\tau ^{1-\alpha }}d\tau =\frac{\Gamma \left( \alpha \right) }{\omega
^{\alpha}}\sin \left( \frac{\alpha \pi }{2}-\omega \right) .
\label{B_inf_resultado}
\end{equation}%
{On the other hand, expanding the sine function in its Maclaurin series (\cite{Olver}, Equation (4.19.1)):}%
\begin{equation}
\sin z=\sum_{k=0}^{\infty }\frac{\left( -1\right) ^{k}\,z^{2k+1}}{\left(
2k+1\right) !},
\end{equation}%
and taking into account the definition of the beta function (\ref{Beta_def}%
), we obtain%
\begin{eqnarray}
B_{1} &=&-\omega \sum_{k=0}^{\infty }\frac{\left( -1\right) ^{k}\omega
^{2k+1}}{\left( 2k+1\right) !}\int_{0}^{1}\tau ^{\alpha -1}\left( 1-\tau
\right) ^{2k+1}d\tau \\
&=&-\omega \sum_{k=0}^{\infty }\frac{\left( -1\right) ^{k}\omega
^{2k}\,\Gamma \left( \alpha \right) }{\Gamma \left( \alpha +2k+2\right) }.
\end{eqnarray}%
According to the definition of the Pochhammer symbol (\ref{pochhammer_def}),
and taking into account the following formulas (\cite{Atlas}, Equations (18:5:4) and (18:5:7)):%
\begin{eqnarray}
\left( x\right) _{2n} &=&4^{n}\left( \frac{x}{2}\right) _{n}\left( \frac{1+x%
}{2}\right) _{n}, \\
\left( x\right) _{n+1} &=&x\,\left( x+1\right) _{n},
\end{eqnarray}%
we have%
\begin{eqnarray}
\frac{\Gamma \left( \alpha +2k+2\right) \,}{\Gamma \left( \alpha \right) }
&=&\left( \alpha \right) _{2\left( k+1\right) }=4^{k+1}\left( \frac{\alpha }{%
2}\right) _{k+1}\left( \frac{1+\alpha }{2}\right) _{k+1} \\
&=&4^{k}\alpha \left( \alpha +1\right) \left( \frac{\alpha }{2}+1\right)
_{k}\left( \frac{3+\alpha }{2}\right) _{k}.
\end{eqnarray}%
Thus,%
\begin{equation}
B_{1}=-\frac{\omega }{\alpha \left( \alpha +1\right) }\sum_{k=0}^{\infty }%
\frac{\,1}{\left( 1+\frac{\alpha }{2}\right) _{k}\left( \frac{3+\alpha }{2}%
\right) _{k}}\left( -\frac{\omega ^{2}}{4}\right) ^{k}.
\end{equation}%
Recalling the definition of the generalized hypergeometric function (\ref{pFq_def}), the last result can be recast as
\begin{equation}
B_{1}=-\frac{\omega }{\alpha \left( \alpha +1\right) }\,_{1}F_{2}\left(
\begin{array}{c}
1 \\
1+\frac{\alpha }{2},\frac{3+\alpha }{2}%
\end{array}%
;-\frac{\omega ^{2}}{4}\right) .  \label{B_1_resultado}
\end{equation}%
Substituting the results (\ref{B_inf_resultado})\ and (\ref{B_1_resultado})\
into (\ref{Ib=B_1+B_inf}), we obtain
\begin{equation}
\int_{0}^{\infty }\frac{\sin \omega t}{\left( 1+t\right) ^{1-\alpha }}\,dt =%
\frac{\Gamma \left( \alpha \right) }{\omega ^{\alpha}}\sin \left( \frac{%
\alpha \pi }{2}-\omega \right) +\frac{\omega }{\alpha \left( \alpha
+1\right) }\,_{1}F_{2}\left(
\begin{array}{c}
1 \\
1+\frac{\alpha }{2},\frac{3+\alpha }{2}%
\end{array}%
;-\frac{\omega ^{2}}{4}\right) .  \label{IB_resultado}
\end{equation}

%%%%%%%%%%%%%%%%%%%%%%%%%%%%%%%%%%%%%%%%%%

%=====================================


\begin{thebibliography}{999}

\bibitem[Bateman and Erd{\'e}lyi(1953)]{Bateman}
Bateman, H.; Erd{\'e}lyi, A.
\newblock {\em Higher {T}ranscendental {F}unctions}; Mc. Graw-Hill Book
  Company:  Columbus, OH, USA, 1953; Volume 3.

\bibitem[Giusti(2017)]{Giusti}
Giusti, A.
\newblock On infinite order differential operators in fractional
  viscoelasticity.
\newblock {\em Fract. Calc. Appl. Anal.} {\bf 2017}, {\em 20},~854--867.


\bibitem[Gonz{\'a}lez-Santander and Mainardi(2024)]{BeckerLomnitzLambert}
Gonz{\'a}lez-Santander, J.L.; Mainardi, F.
\newblock A Comparative view of {B}ecker, {L}omnitz, and {L}ambert linear
  viscoelastic models.
\newblock {\em Mathematics} {\bf 2024}, {\em 12},~3426.
\bibitem[Gradshteyn and Ryzhik(2014)]{Gradshteyn}
Gradshteyn, I.S.; Ryzhik, I.M.
\newblock {\em Table of {I}ntegrals, {S}eries, and {P}roducts}; Academic Press:  Cambridge, MA, USA, 2014.

\bibitem[Gross(1953)]{GrossBOOK53}
Gross, B.
\newblock {\em Mathematical {S}tructure of the {T}heories of
  {V}iscoelasticity}; Hermann \& C.: Paris, France 1953.

\bibitem[Hanyga(2005)]{HanygaRHEOACTA05}
Hanyga, A.
\newblock Viscous dissipation and completely monotonic relaxation moduli.
\newblock {\em Rheol. Acta} {\bf 2005}, {\em 44},~614--621.

\bibitem[Jeffreys(1958)]{JeffreysLomnitz}
Jeffreys, H.
\newblock A modification of {L}omnitz's law of creep in rocks.
\newblock {\em Geophys. J. Int.} {\bf 1958}, {\em 1},~92--95.

\bibitem[Lebedev(1965)]{Lebedev}
Lebedev, N.N.
\newblock {\em Special {F}unctions and Their {A}pplications}; 
Prentice-Hall,  Inc.:  Hoboken, NJ, USA, 1965.

\bibitem[Lomnitz(1957)]{LomnitzQ}
Lomnitz, C.
\newblock Linear dissipation in solids.
\newblock {\em J. Appl. Phys.} {\bf 1957}, {\em 28},~201--205.


\bibitem[Mainardi(2022)]{MainardiBook}
Mainardi, F.
\newblock {\em Fractional {C}alculus and {W}aves in {L}inear {V}iscoelasticity:
  An {I}ntroduction to {M}athematical {M}odels}; World Scientific: Singapore %MDPI: Please add the correct publisher location (city, conuntry).
   2022.

\bibitem[Mainardi et~al.(2023)]{LambertJL}
Mainardi, F.; Masina, E.; Gonz{\'a}lez-Santander, J.L.
\newblock A note on the {L}ambert {W} function: {B}ernstein and {S}tieltjes
  properties for a creep model in linear viscoelasticity.
\newblock {\em Symmetry} {\bf 2023}, {\em 15},~1654.

\bibitem[Mainardi et~al.(2019)]{MainardiBecker}
Mainardi, F.; Masina, E.; Spada, G.
\newblock A generalization of the {B}ecker model in linear viscoelasticity:
  Creep, relaxation and internal friction.
\newblock {\em Mech. Time-Depend. Mat.} {\bf 2019}, {\em 23},~283--294.

\bibitem[Mainardi and Spada(2012)]{MainardiLomnitz}
Mainardi, F.; Spada, G.
\newblock On the viscoelastic characterization of the Jeffreys-Lomnitz law
  of creep.
\newblock {\em Rheol. Acta} {\bf 2012}, {\em 51},~783--791.

\bibitem[Oldham et~al.(2009)]{Atlas}
Oldham, K.B.; Myland, J.; Spanier, J.
\newblock {\em An {A}tlas of {F}unctions: With {E}quator, the {A}tlas
  {F}unction {C}alculator}; Springer: New York, NY, USA, 2009.
  
  \bibitem[Olver et~al.(2010)]{Olver}
Olver, F.W.J.; Olde~Daalhuis, A.B.; Lozier, D.W.; Schneider, B.I.; Boisvert,
  R.F.; Clark, C.W.; Miller, B.R.; Saunders, B.V.; Cohl, H.S.; McClain, M.A.
  (Eds.)
\newblock {\em {NIST Handbook of Mathematical Functions}};
 Cambridge University  Press: New York, NY, US

\bibitem[Pipkin(1986)]{PipkinBOOK86}
Pipkin, A.C.
\newblock {\em Lectures on {V}iscoelasticity {T}heory}; Springer Science \&
  Business Media: New York, NY, USA, 1986.
  
\bibitem[Prudnikov et~al.(1986)]{Prudnikov3}
Prudnikov, A.P.; Brychkov, Y.A.; Marichev, O.I.
\newblock {\em Integrals and {S}eries: {M}ore {S}pecial {F}unctions};
  CRC Press:  Boca Raton, FL, USA,
1986; Volume 3.

\bibitem[Prudnikov et~al.(1986)]{Prudnikov4}
Prudnikov, A.P.; Brychkov, Y.A.; Marichev, O.I.
\newblock {\em Integrals and {S}eries: {D}irect {L}aplace {T}ransforms};
 CRC Press:  Boca Raton, FL, USA,
1986;  Volume 4.

\bibitem[Schiff(1999)]{Schiff}
Schiff, J.L.
\newblock {\em The {L}aplace {T}ransform: {T}heory and {A}pplications};
 Springer Science \& Business Media: New York, NY, USA, 1999.


\bibitem[Strick and Mainardi (1982)]{StrickMainardi}
Strick, E.; Mainardi, F.
\newblock On a general class of constant $Q$ solids.
\newblock {\em Geophys. J. Royal Astr. Soc.} {\bf 1982}, {\em 69},~415--429.

\bibitem[Tschoegl(1989)]{TschoegelBOOK89}
Tschoegl, N.W.
\newblock {\em The {P}henomenological {T}heory of {L}inear {V}iscoelastic
  {B}ehavior: An {I}ntroduction}; Springer: New York, NY, USA, 1989.


\end{thebibliography}
\end{document}